\documentclass[aps,twocolumn,amsmath,showpacs,amssymb,
superscriptaddress,nofootinbib]{revtex4-1}

\usepackage{graphics,graphicx}
\usepackage{amsmath, amssymb}
\usepackage{multirow}
\usepackage{longtable}
\usepackage{color}
\usepackage{hyperref}

\usepackage[normalem]{ulem}  

\begin{document}

\title{ Thermal contribution of unstable states }
\date{\today}

\author{Pok Man Lo}
\affiliation{Institute of Theoretical Physics, University of Wroclaw,
PL-50204 Wroc\l aw, Poland}
\author{Francesco Giacosa}
\affiliation{Institute of Physics, Jan-Kochanowski University, 25-406 Kielce, Poland}

\begin{abstract}

  Within the framework of the Lee model, we analyze in detail the difference between the 
  energy derivative of the phase shift and the standard spectral function of the unstable state.
  The fact that the model is exactly solvable allows us to demonstrate the construction of these observables
  from various exact Green functions. The connection to a formula due to Krein, Friedal, and Lloyd is also examined.
  We also directly demonstrate how the derivative of the phase shift correctly identifies 
  the relevant interaction contributions
  for consistently including an unstable state in describing the thermodynamics.

\end{abstract}

\maketitle

\section{Introduction}
\label{sec1}

Formal treatment of the interactions in a gas of particles at finite temperature is an 
important topic in thermal field theory~\cite{bloch,campa,satz}. 
In particular, a consistent description of the unstable states is imperative for understanding the hadron gas, 
see, e.g., Refs.~\cite{satz,wfbook,dmb,venugopalan}. 
Some questions of interest include:
How are the bulk properties of the medium, such as pressure and energy density, affected by unstable particles? 
What are the effective ways to take these into consideration? 
What insights can be gained from comparing different approaches, e.g., 
the standard imaginary time formalism 
and those based on a virial expansion?~\cite{dmb,osborn,fre,chaichian,LeClair:2006hi,how1,virial,smat}
In this work we tackle these issues using an effective Hamiltonian approach.
Besides an intuitive modification driven by the spectral function of the unstable particle (the resonant contribution),
we shall demonstrate how the very presence of an interaction modifies the 2-body states composed of the stable particles (the nonresonant contribution).
The sum of these modifications recovers a well-known result~\cite{bu,dmb,weinhold}, according to which 
the derivative of the scattering phase shift can be identified as the density of state for computing the partition function.

Many thermal models have the widths of the resonances implemented but the nonresonant interactions are neglected.
This can lead to misleading results when interpreting the contribution from an interaction channel~\cite{sigma,kappa}. 
An illustrative example is the pion-pion scattering in the $I=0$ channel, 
in which the famous $f_0(500)$ resonance, a.k.a. the $\sigma$-meson, is involved. 
The empirical phase shift, analyzed by the chiral perturbation theory~\cite{Oller:1998hw}, 
reveals that there are substantial (effective) repulsive corrections coming from the exchange interactions in the t- and u-channels. 
Additional cancellation also comes from the $I=2$ channel.
A model that corrects only for the width of the resonance is incapable of handling these effects.
Some models try to remedy this by introducing extraneous repulsive forces, e.g., via an excluded volume. 
This, however, will generally lead to a model which contradicts the known phase shift~\cite{exclvol}.

In this study we consider a system of stable particles ``$\pi$'' (the pions) and 
unstable particle ``$\rho$'' (the $\rho$-mesons).~\footnote{ 
These notions obviously mirror the case of a thermal hadron gas with pions and $\rho(770)$'s. 
However, our discussion generally applies to any thermal system with unstable states.}
Each $\rho$ can decay into two $\pi$'s via the interaction $\rho \rightarrow \pi\pi$. 
Chains of interactions, e.g., $\pi\pi \rightarrow \rho \rightarrow \pi\pi$, are also included.
The fact that only two types of particles are considered drastically simplifies the discussion, 
but the main non-trivial features are kept.

We use a Lee-Model Hamiltonian (LH)~\cite{TDLEE,Chiu:1992pc} to describe the interactions in the $\rho \pi$ system.
Similar Hamiltonians have been used to explore various areas in physics, ranging 
from atomic physics and quantum optics~\cite{Jaynes:1963zz,Babelon:2007td, Babelon:2009dra,scully} to baryon decays \cite{Liu:2015ktc}. 
The Lee model offers a useful theoretical setup to study an interacting system~\cite{zeno,Facchi:2000,Facchi:1999ik,Facchi:1999nq,Berman,Kofman,fr1} 
and in many aspects it resembles a quantum field theory~\cite{fr2}. 
In practice, the LH contains an unstable state $\left\vert \rho\right\rangle $, 
as well as a continuum of 2-body states $ \left\vert \pi(\vec{q})\pi(-\vec{q})\right\rangle $ 
with all possible relative momenta $\vec{q}$ of the pair. 
The decay $\rho\rightarrow\pi\pi$ is described by mixing terms of the form:
$\sim g \left( \left\vert\rho\right\rangle \left\langle \pi\pi\right\vert +\left\vert \pi 
\pi\right\rangle \left\langle \rho\right\vert \right)$. 
This gives the unstable particle $\rho$ a width, i.e., a distribution in energies (or masses) dictated 
by the spectral function $A_\rho(E)$~\cite{fr2,salam1,salam2,fr4}, such that $A_\rho(E)\, \frac{dE}{2\pi}$ can be 
interpreted as the probability that the unstable $\rho$ has energy between $E$ and $E+dE$.
The spectral function can be calculated from the imaginary part of the $\rho$-propagator. See Sec.~\ref{sec3} for details.
For a narrow-width state this can be approximated by the Breit-Wigner (BW) formula~\cite{bw1,bw2}, 

\begin{align}
  \label{eq:bw}
  A_\rho(E) \approx A_\rho^{BW}(E) = \frac{ \Gamma_{\rho\pi\pi} }{ (E-m_\rho)^2 + \frac{\Gamma_{\rho\pi\pi}^2}{4}}. 
\end{align}

\noindent Note that the width $\Gamma_{\rho\pi\pi}$ is generally energy dependent, 
and the mass $m_\rho$ can be modified by the real part of quantum loops.
Thus, the decay probability is never exactly exponential. 
See, e.g., the theoretical treatment in Refs.~\cite{zeno,Fonda:1978dk,ersak} and 
the experimental results in Refs.~\cite{ex1,ex2,ex3}; 
and Ref.~\cite{fr3} for a discussion in a quantum field theory.

Based on the LH, the finite temperature properties of the system 
can be derived using the standard techniques of statistical mechanics. See Sec.~\ref{sec4} for details.
Here we give a synopsis of our discussion. 
Consider the hypothetical limit where $\rho$ and $\pi$'s are completely decoupled, i.e., $g = 0$ (or $\Gamma_{\rho\pi\pi}=0$). 
The pressure of the system would be given by a sum of two contributions: 

\begin{align}
  \label{eq:p0}
  P_{\rm free}=P^{(0)}_{\pi}+P^{(0)}_\rho,
\end{align}
  
\noindent where P$^{(0)}_{a}=P^{(0)}(E=m_a,T)$ denotes the pressure of an ideal gas of a species $a$, which depends on its mass ($m_a$) and degeneracy. The dependence on temperature $T$ is understood.
Even if the interaction is switched on, Eq.~\eqref{eq:p0} can still be used as an estimate of the pressure for a narrow-width $\rho$. 
This is the fundamental premise of the hadron resonance gas (HRG) model~\cite{Hagedorn:1965st,hrgnature}: contribution of resonances to the thermodynamics is approximated by an uncorrelated gas of zero-width particles.

As a next step in improving the approximation, we take into account the width of $\rho$ via a weighed sum by $A_\rho(E)$:

\begin{align}
  P^{(0)}_\rho \rightarrow P_\rho = \int \frac{dE}{2\pi} \, A_\rho(E) \times P^{(0)}(E,T)
\end{align}

\noindent such that the total pressure is approximated as (Scheme-A):

\begin{align}
  \label{eq:pA}
  P_{\rm Sch. A} = P^{(0)}_{\pi}+P_\rho. 
\end{align}

\noindent This scheme is employed in many versions of the HRG models, see e.g., Refs.~\cite{satz,wfbook,hrg1,hrg2,hrg3,hrg4}. 
See also the K-matrix-based approach~\cite{Doring:2006ue,kmat}.
However, Eq.~\eqref{eq:pA} is not yet complete. 
According to the S-matrix formulation of statistical mechanics by Dashen {\it et al.}~\cite{dmb} (see also the discussion by Weinhold {\it et al.}~\cite{weinhold}), 
the correct result of the pressure at arbitrary $g$ (or $\Gamma_{\rho\pi\pi}$) is given by (S-matrix scheme): 


\begin{align}
  \label{eq:pB}
  \begin{split}
  P_{\rm S-matrix} &= P^{(0)}_{\pi} +  \int \frac{dE}{2\pi} \, B(E) \times P^{(0)}(E,T) \\
  B(E) &= 2 \, \frac{d}{d E} {\mathcal{Q}}_{\pi\pi}(E). 
  \end{split}
\end{align}

\noindent where ${\mathcal{Q}}_{\pi\pi}(E)$ is the phase shift for the scattering process $\pi \pi \rightarrow \pi \pi$. 
Here we summarize some key features of the S-matrix scheme:

\begin{itemize}

  \item[(i)] Observe that there is no explicit $\rho$ contribution in Eq.~\eqref{eq:pB}:
The pressure is determined based on the scattering information of the asymptotic (stable) states alone.
In fact, it is not compulsory to introduce the $\rho$ state as an explicit degree of freedom. 
Its presence is encoded in the phase shift. This point will be made clear by direct model calculations.

  \item[(ii)] Eqs.~\eqref{eq:pA} and~\eqref{eq:pB} reduce to the free gas result~\eqref{eq:p0} in the limit of 
$g \rightarrow 0$ ( or $\Gamma_{\rho\pi\pi} \rightarrow 0$ ).~\footnote{The $g=0$ limits of $A_\rho(E)$ and $B(E)$ are not well-defined.}
For the latter, we have

\begin{align}
  \begin{split}
  \delta_{\pi\pi}(E) &\rightarrow \pi \, \theta(E-m_\rho) \\
    B(E) &\rightarrow 2 \pi \, \delta(E-m_\rho). 
  \end{split}
\end{align}

\item[(iii)] Generally, $B(E) \neq A_\rho(E)$, and Eqs.~\eqref{eq:pA} and ~\eqref{eq:pB} are thus different. 
Systems which show substantial deviation are plenty:
In addition to the case of $f_0(500)$ mentioned, 
nonresonant contribution is found to be important in the study of $\kappa(700)$~\cite{sigma,kappa}, 
the $N^*$ and $\Delta$ resonances~\cite{chibq,ppuzz}, the $S=-1$ hyperons~\cite{chibs}, etc. 
It is also the case for the recently discovered $X,$ $Y,$ $Z$ states~\cite{exotica,xyz1,xyz2}. 
As shown in a recent work~\cite{Ortega:2017hpw}, the state $X(3872)$ makes only a small contribution to the thermodynamics due to nonresonant effects. 
For what concerns other states, future studies based on Eq.~\eqref{eq:pB} are needed.

\end{itemize}

In this work we verify Eq.~\eqref{eq:pB}, instead of Eq.~\eqref{eq:pA}, 
gives the correct description of the thermodynamics of an interacting system. 
This point had been raised in previous works, see e.g., Ref.~\cite{weinhold}, 
but the actual adoption of the scheme remains limited~\cite{Broniowski_2003,ppuzz}.
We hope that a more detailed account of the different spectral functions 
can raise the awareness of the issue in the community and further promote the use of the correct formula.
We also establish their formal relations to the resolvent and the density of states. 
This gives an interesting perspective in describing the thermodynamics of an interacting system.

Using an LH, we derive the mismatch analytically, 
the result takes the form

\begin{align}
  \label{eq:gd}
    B(E) = A_\rho(E) + \sum_q \Delta A_{2 \pi}(E;q),
\end{align}

\noindent where the second term in the R.H.S. describes the 
the modification of the spectral function of the 2-body $\vert\pi\pi\rangle$ state. 
Such a term is present even in the absence of the resonance, e.g., taking the large $m_\rho$ limit. (See Sec.~\ref{sec3}) 
The correct expression of the pressure can be decomposed as:

\begin{align}
  P_{\rm S-matrix} = P^{(0)}_{\pi} +  P_{\rho} +  \Delta P_{2\pi}, 
\end{align}

\noindent where

\begin{align}
  \Delta P_{2 \pi} = \int \frac{dE}{2\pi} \, \sum_q \Delta A_{2 \pi}(E;q) \times P^{(0)}(E,T).
\end{align}

\noindent As we shall see, the last term is in general not negligible and can even be dominant at low temperatures.

The paper is organized as follows: In Sec.~\ref{sec2} the details of the Lee model are
presented. Then, in Sec.~\ref{sec3}, the spectral functions for both $\rho$ and $\pi\pi$, 
and the phase shift are introduced. Here the important Eq.~\eqref{eq:gd} is derived. 
In Sec.~\ref{sec4} the thermodynamic properties of the system are determined
analytically, with special focus on the pressure with its various contributions.
A numerical example shows that $\Delta P_{2\pi}$ can be sizable and in general should not be neglected. 
Finally, discussions and conclusions are given in Sec.~\ref{sec5}.

\section{The Lee Model}
\label{sec2}

The Lee model~\cite{TDLEE} describing the $\rho \leftrightarrow \pi \pi$ system can be formulated as follows.~\footnote{
In the following we measure energy with respect to $2 m_\pi$, 
and the nonrelativistic dispersion relation $\epsilon(q) = {q^2}/{m_\pi}$ is used.
We also choose to present our model in a discretized form.
This prepares for the later numerical treatment of solving the system on a momentum grid.~\cite{FVH}. 
It is easy to go to the continuum by taking $ \sum_q C(q)^2 \, (\cdots) \rightarrow \int \frac{d^3 q}{(2 \pi)^3} \, (\cdots)$.
}
Introducing the basis states in the center-of-mass (CM) frame:

\begin{align}
  \vert \rho \rangle, \{ \vert q \rangle \},
\end{align}

\noindent where $q$ is the momentum label for the two-pion state $\vert \pi(\vec{q}) \pi(-\vec{q}) \rangle$. 
The Hamiltonian of the system can be represented as an $(1+N_q)\times(1+N_q)$-matrix. 
The non-interacting Hamiltonian $H_0$ is a diagonal matrix with

\begin{align}
  \begin{split}
    H_0 \vert \rho \rangle = \Delta_0 \vert \rho \rangle, \\
    H_0 \vert q \rangle = \epsilon(q) \vert q \rangle.
  \end{split}
\end{align}


The interaction $V$ describes the coupling of $\rho$ with the $\vert q \rangle$ states

\begin{align}
  \label{eq:vint}
    V = \sum_q g_{\rm eff} \,  C(q) \left ( \vert \rho \rangle  \langle q \rangle + \vert q \rangle  \langle \rho \rangle \right), 
\end{align}

\noindent such that

\begin{align}
  \begin{split}
    V \vert \rho \rangle &= \sum_{q^\prime}  g_{\rm eff} \, C(q^\prime) \vert q^\prime \rangle  \\
    V \vert q \rangle &= g_{\rm eff} \, C(q) \vert \rho \rangle.
  \end{split}
\end{align}

\noindent We use

\begin{align}
  \begin{split}
C(q)^2 = \frac{4 \pi q^2}{(2 \pi)^3} \, \delta q
  \end{split}
\end{align}

\noindent to implement the spherical degeneracy. Note that $\delta q = \frac{2 \pi}{L}$ in the finite volume formulation, $L$ being the size of the box. 
The coupling $g_{\rm eff}$ is generally $q$-dependent. The full Hamiltonian then reads

\begin{align}
  H = H_0 + V.
\end{align}

With the Hamiltonian defined, we can construct the resolvent operators

\begin{align}
  \begin{split}
    G_0(E) &= \frac{1}{E-H_0 + i \epsilon} \\
    G(E) &= \frac{1}{E-H + i \epsilon},
  \end{split}
\end{align}

\noindent which can be understood again as $(1+N_q)\times(1+N_q)$ matrices. 
In the remainder of this paper we will suppress the $E$ dependence unless there is a chance for confusion.

The well-known relations from the Lippmann-Schwinger equation can also be directly realized:

\begin{align}
  \begin{split}
    G &= G_0 + G_0 V G \\
    &= G_0 + G_0 T G_0,
  \end{split}
\end{align}

\noindent and

\begin{align}
  \begin{split}
    T & = V + V G_0 T \\
    &= V + V G V \\
    &= V G G_0^{-1}.
  \end{split}
\end{align}

In this paper, we investigate the inclusion of an unstable state
in the description of thermodynamics using the S-matrix formulation.
The key operators of interest in this scheme is the scattering operator~\cite{smat, taylor}

\begin{align}
  \begin{split}
    \hat{S} &= G^*_0 \, {G^*}^{-1} \, G \, {G_0}^{-1} \\
    &= I + \left( G_0-G^*_0 \right) \times V G G_0^{-1} \\
    &= I - 2 \pi i \, \delta(E-H_0) \times T.
  \end{split}
\end{align}

\noindent Since $\rho$ is not an asymptotic state, the actual scattering matrix $S$ is extracted from the (lower-right) $N_q \times N_q$ block of $\hat{S}$. 
In addition, we introduce an operator $\mathcal{K}$, due to Krein, Friedal and Lloyd (KFL)~\cite{chaos,rev1}, 
defined as the difference of the spectral functions 

\begin{align}
  \label{eq:kfl}
  \begin{split}
    \mathcal{K} &= -{\rm Im } \, \left[ (G-G^*)-(G_0-G_0^*) \right] \\
                &= A - A_0 \\
                &= \Delta A,
  \end{split}
\end{align}

\noindent where we have identified the spectral function operator

\begin{align}
  \label{eq:afunc}
  \begin{split}
    A(E) & = -2 \, {\rm Im} \, G  = -{\rm Im} \, ( G-G^* )  \\
    A_0(E) & = -2 \, {\rm Im} \, G_0 = -{\rm Im} \, (G_0-G_0^*).
  \end{split}
\end{align}

\noindent These operators are deeply connected with the scattering phase shift $\mathcal{Q}$ 
and the effective spectral function $B$. The latter is defined as

\begin{align}
  \label{eq:bfunc}
  B(E) = 2 \frac{\partial}{\partial E} \mathcal{Q}(E).
\end{align}

\noindent There are multiple ways to extract the phase shift $\mathcal{Q}$ from the resolvents (or Green functions). 
(See Eqs.~\eqref{eq:ps1},~\eqref{eq:recipe}). 
These and the explicit relations among the various observables 
will be demonstrated in the context of the Lee model in Sec.~\ref{sec3}.

\section{Phase shift and effective spectral function}
\label{sec3}

In the Lee model various theoretical quantities, e.g. the propagator and the self-energy of $\rho$, 
can be analytically computed. It is a useful exercise to revisit these formulas as it helps
to build an understanding the physical content of the KFL operator $\mathcal{K}$ and the derivative of the phase shift.

\subsection{The $\rho$-propagator}

The $\rho$-propagator $G_\rho$ can be computed from $\langle \rho \vert G(E) \vert \rho \rangle$, i.e. the first diagonal entry of the matrix $G(E)$, as

\begin{align}
  \begin{split}
  G_\rho &\equiv \langle \rho \vert G \vert \rho \rangle \\ 
  & = \langle \rho \vert G_0 \vert \rho \rangle  + \langle \rho \vert G_0 V G_0 \vert \rho \rangle  +
\langle \rho \vert G_0 V G_0 V G_0 \vert \rho \rangle + \ldots,
  \end{split}
\end{align}

\noindent where each term can be directly worked out:

\begin{align}
  \begin{split}
    \langle \rho \vert G_0 \vert \rho \rangle &= G^0_\rho = \frac{1}{E-\Delta_0+i \epsilon}  \\
    \langle \rho \vert G_0 V G_0 \vert \rho \rangle &= 0, 
  \end{split}
\end{align}

\noindent and

\begin{align}
  \langle \rho \vert G_0 V G_0 V G_0 \vert \rho \rangle &=  G^0_\rho \left( \sum_q C(q)^2 g_{\rm eff}^2 G^0_{2 \pi}(E;q) \right) G^0_\rho.
\end{align}

\noindent It is clear that the propagator $G_\rho$ can be re-summed to all orders via

\begin{align}
  \begin{split}
    G_\rho &= G^0_\rho +  G^0_\rho \Sigma_\rho G^0_\rho  + G^0_\rho  \Sigma_\rho G^0_\rho  \Sigma_\rho G^0_\rho + \ldots \\
    &= \frac{1}{(G^0_\rho)^{-1}-\Sigma_\rho},
  \end{split}
\end{align}

\noindent where the self-energy of $\rho$ can be explicitly computed by

\begin{align}
\label{eq:sigma}
  \begin{split}
    \Sigma_\rho &\equiv  \langle \rho \vert V G_0 V \vert \rho \rangle \\
    &= \sum_q C(q)^2 g_{\rm eff}^2 G^0_{2 \pi}(E;q).
  \end{split}
\end{align}

A relation that will prove useful later is the energy-derivative of $\Sigma_\rho$:

\begin{align}
  \begin{split}
    \frac{\partial}{\partial E}\Sigma_\rho &=  -1 \times \sum_q C(q)^2 g_{\rm eff}^2 G^0_{2 \pi}(E;q) G^0_{2 \pi}(E;q).
  \end{split}
\end{align}

\noindent which is easily seen by noting 

\begin{align}
  \begin{split}
    G^0_{2 \pi}(E;q) &= \frac{1}{E-\epsilon(q)+i \epsilon} \\
    \implies \frac{\partial}{\partial E} G^0_{2 \pi}(E;q) &= - G^0_{2 \pi}(E;q) \times G^0_{2 \pi}(E;q)\\
  \end{split}
\end{align}

\subsection{The 2-pion propagator}

Now we turn to the 2-pion states. The propagator $G_{2 \pi}(E;q)$ can be explicitly worked out from the corresponding diagonal entries of $G(E)$:

\begin{align}
  \begin{split}
    G_{2 \pi}(E;q) &= \langle q \vert G \vert q \rangle \\ 
    &= G^0_{2 \pi}(E;q)  + \langle q \vert G_0 T G_0 \vert q \rangle \\ 
    &= G^0_{2 \pi}(E;q)  + G^0_{2 \pi}(E;q) \times \langle q \vert T \vert q \rangle \times G^0_{2 \pi}(E;q).      
  \end{split}
\end{align}

\noindent We now show that the diagonal T-matrix is directly related to the full propagator $G_\rho$. 
This is an important relation, as it dictates 
how the properties of the unstable state can be inferred from the scattering of the stable particles.
To see that we employ the following expression of the T-matrix:

\begin{align}
    \langle q \vert T \vert q \rangle &= \langle q \vert V \vert q \rangle + \langle q \vert VGV \vert q \rangle.
\end{align}

\noindent The first term is $0$ since $V$ is off-diagonal. The second term gives

\begin{align}
  \langle q \vert VGV \vert q \rangle = g_{\rm eff}^2 C(q)^2 \times  G_\rho
\end{align}

\noindent and hence

\begin{align}
    \label{eq:tmat}
    \langle q \vert T \vert q \rangle &=  g_{\rm eff}^2 C(q)^2 \times G_\rho.
\end{align}

\noindent This relates the amplitude of $\pi \pi$ scatterings to the $\rho$-propagator. 
Indeed, in the simple setting of the Lee model, all the physical information concerning the unstable state $\rho$ 
can be extracted from the diagonal T-matrix $ \langle q \vert T \vert q \rangle $. 

Finally, the full propagator $G_{2 \pi}(E;q)$ can be obtained in closed form as

\begin{align}
  G_{2 \pi}(E;q) &= G^0_{2 \pi}(E;q)  + G^0_{2 \pi}(E;q) \left( C(q)^2 g_{\rm eff}^2 G_\rho \right) G^0_{2 \pi}(E;q).
\end{align}

\begin{figure*}[ht!]
\centering
\includegraphics[width=0.49\textwidth]{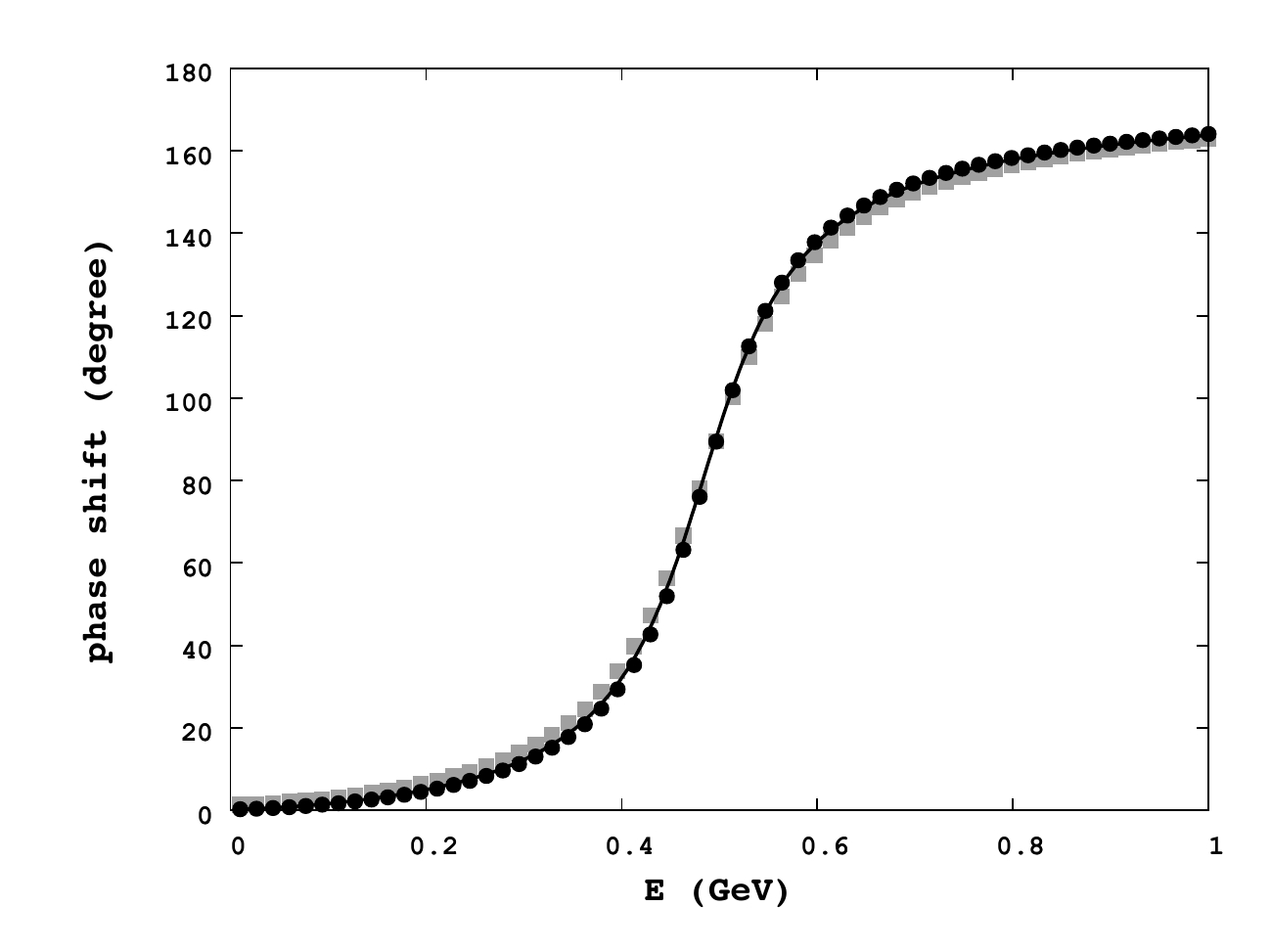}
\includegraphics[width=0.49\textwidth]{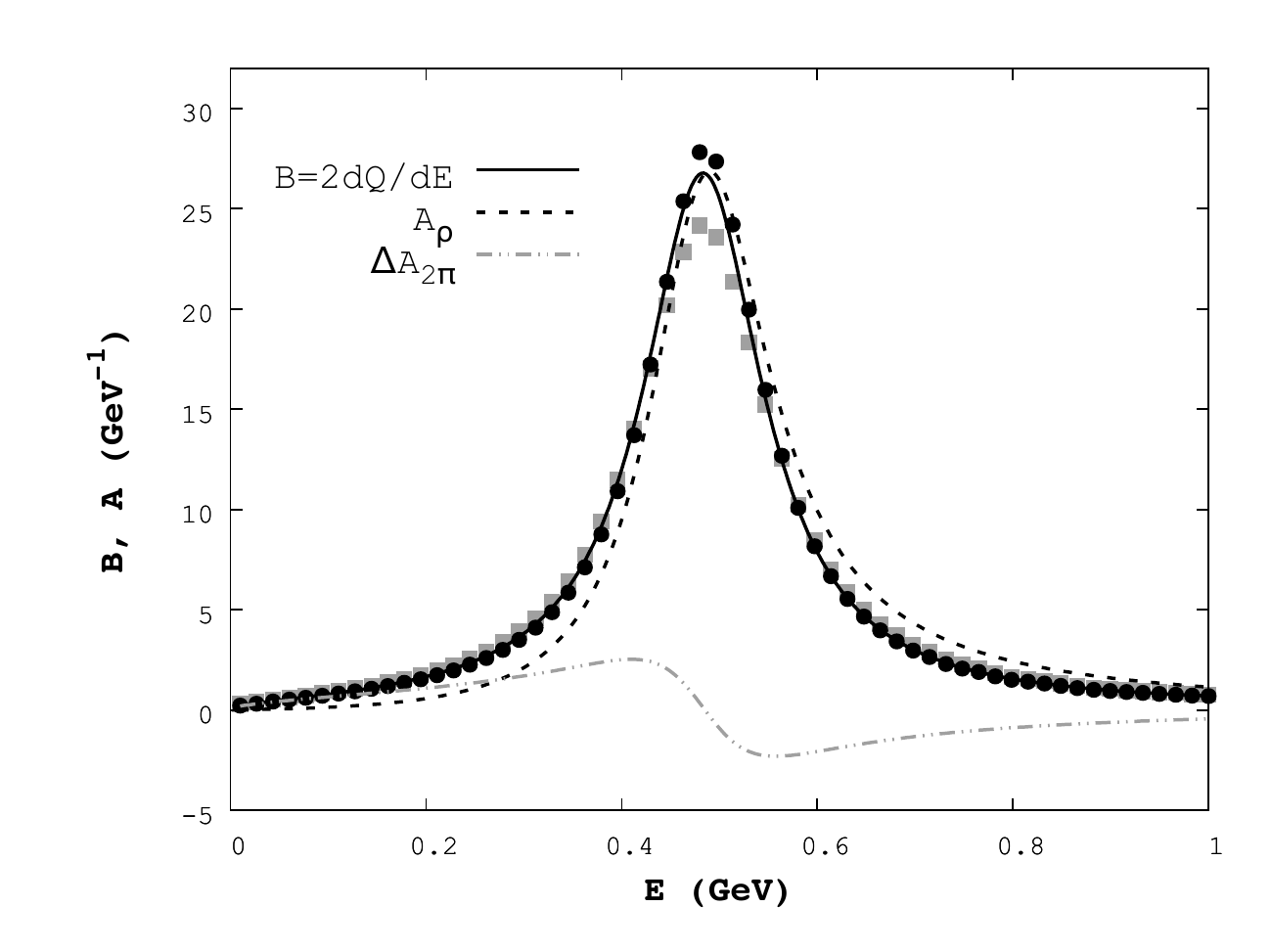}
  \caption{Phase shift computed from the Lee model (Eq.~\eqref{eq:ps1}) and the corresponding effective spectral functions (Eq.~\eqref{eq:afunc}, ~\eqref{eq:bfunc}) for the $\rho$-meson. 
  $E$ is the energy of the relative motion. Note that $B = A_\rho + \Delta A_{2 \pi}$. 
  The points are the numerical results obtained from the momentum grid method:
  Grey squares indicate results based on Eq.~\eqref{eq:ps1}, while 
  black circles indicate results based on Eq.~\eqref{eq:recipe}.
  See text.
  }
\label{fig:one}
\end{figure*}

\subsection{Effective spectral function}

We are now ready to examine the expressions of the phase shift $\mathcal{Q}$, the effective spectral function $B$ 
and the operator $\mathcal{K}$ in the context of the Lee model.

The phase shift can most simply be extracted from $G_\rho$ via~\footnote{One can obtain the phase shift directly from the S-matrix. See Eq.~\eqref{eq:recipe}.}

\begin{align}
  \label{eq:ps1}
  \begin{split}
    \mathcal{Q}(E) &= {\rm Im} \, \ln G_\rho \\
    &= \tan^{-1} \, \frac{{\rm Im} \, \Sigma_\rho}{E-\Delta_0-{\rm Re} \Sigma_\rho}.
  \end{split}
\end{align}

\noindent The effective spectral function $B$ is known to be related to the interacting part of the density of state.
In the Lee model, it is 

\begin{align}
  \begin{split}
    B(E) &= 2 \frac{\partial}{\partial E} \mathcal{Q}(E) \\
    &= 2 \frac{\partial}{\partial E} {\rm Im} \, \ln G_\rho \\
    &= -2 \, {\rm Im} \, \left( G_\rho \frac{\partial}{\partial E} G_\rho^{-1} \right)\\
    &= -2 \, {\rm Im} \, \left[ G_\rho \frac{\partial}{\partial E} ( E - \Delta_0 - \Sigma_\rho) \right] \\
    &= -2 \, {\rm Im} \, G_\rho  +  2 \, {\rm Im} \, \left( G_\rho \times \frac{\partial}{\partial E}\Sigma_\rho \right),
  \end{split}
\end{align}

\noindent using the relation previously obtained

\begin{align}
  \begin{split}
    \frac{\partial}{\partial E}\Sigma_\rho &=  -1 \times \sum_q C(q)^2 g_{\rm eff}^2 G^0_{2 \pi}(E;q) G^0_{2 \pi}(E;q), 
  \end{split}
\end{align}

\noindent we get 

\begin{align}
  \begin{split}
    B &= -2 \, {\rm Im} \, G_\rho  +  2 \, {\rm Im} \, \left( G_\rho \times \frac{\partial}{\partial E}\Sigma_\rho \right)\\
    &= -2 \, {\rm Im} \, G_\rho  - 2 \, {\rm Im} \, \left[ G_\rho \sum_q  C(q)^2 g_{\rm eff}^2 G^0_{2 \pi}(E;q) G^0_{2 \pi}(E;q) \right]\\
    &= -2 \, {\rm Im} \, G_\rho  - 2 \, {\rm Im} \, \sum_q  \left( G_{2 \pi}(E;q) - G^0_{2 \pi}(E;q) \right).
  \end{split}
\end{align}

\noindent Comparing with the expression of the KFL operator $\mathcal{K}$ in Eq.~\eqref{eq:kfl}, we obtain

\begin{align}
  \begin{split}
    \label{eq:golden}
    B(E) &= A_\rho(E) + \sum_q \Delta A_{2 \pi}(E;q) \\
    &= A^0_\rho(E) + {\rm tr} \mathcal{K} \\
    &= A^0_\rho(E) + \left( \Delta A_{\rho}(E)  + \sum_q \Delta A_{2 \pi}(E;q) \right).
  \end{split}
\end{align}

Relation~\eqref{eq:golden} constitutes the main result of this work. 
It demonstrates how the operator $B$ extracts the physical content, including the contribution from the unstable state $\rho$, of the system. From the first line of Eq.~\eqref{eq:golden}, we see that $B$ includes the contribution from the full spectral function $A_\rho$ and the 2-pion nonresonant interaction $\sum_q  \Delta A_{2 \pi}(E;q) $. 
The second line offers an alternative, but equivalent interpretation: $B$ includes the contribution from the bare-$\rho$, together with the interaction contribution contained in ${\rm tr} \, \mathcal{K}$. The latter includes contributions from the change in the energy spectra of both the $\rho$ and the 2-pion states. 

To understand relation~\eqref{eq:golden} better, we consider the interesting limit of vanishing coupling $g_{\rm eff} \rightarrow 0$. At this limit, the KFL operator $\mathcal{K}$ vanishes by definition. However the phase shift derivative operator $B$ 
would give 

\begin{align}
  B \rightarrow A^0_\rho, 
\end{align}

\noindent that is, it becomes a Dirac-delta function for the bare $\rho$ state. 
It follows that the phase shift $\mathcal{Q}$ 
would becomes a step function $\mathcal{Q} \rightarrow \pi \times \theta(E-\Delta_0)$. 
This is an intuitive limit for describing a $\rho$ that decouples from the pions: 
$\rho$ becomes a stable state, its width ceases to exist
and the state should be included as particles in the asymptotic state. 
These are automatically implemented when the effective spectral function $B$ is used.

Another interesting limit is that of large bare resonant mass $\Delta_0 >> E$. 
In this case, the resonant structure is suppressed, and the nonresonant term dominates.
One can show that as $E \rightarrow 0$

\begin{align}
  \begin{split}
    B &\approx 2 \, ({\rm Re} \,  G_\rho) \times \frac{\partial}{\partial E} {\rm Im} (\Sigma_\rho) \\
    &\approx 2 \, a_l \times \frac{\partial}{\partial E}  q^{2l+1}.
  \end{split}
\end{align}

\noindent where $a_l$ is the scattering length of the channel and $l$ is the relative orbital angular momentum between the pions. Note that terms that are proportional to ${\rm Im} (\Sigma_\rho)$ are subleading relative to $\frac{\partial}{\partial E}{\rm Im} (\Sigma_\rho)$, as the latter is of $\mathcal{O}(q^{2l-1})$.
This should be distinguished from the residual effect of the resonance width at threshold, which is 
of $\mathcal{O}(q^{2l+1})$. Thus, even an energy dependent Breit-Wigner model can not capture the effect of this term.
In addition, the scattering lengths are well constrained by the chiral perturbation theory, 
and indeed the stated form of $B$ was derived~\cite{chiptvirial}.


\subsection{Numerical results}

As a numerical exercise, we solve the Lee Model on a momentum grid following the method of Ref.~\cite{FVH}.
The Hamiltonians are constructed directly as an $(1+N_q)\times(1+N_q)$ matrix. 
The various Green functions are computed by matrix inversions.  
The method is very robust, and the procedure is as follows:

\begin{itemize}

  \item[(i)] Construct the matrices $H$ and $H_0$.

  \item[(ii)] For each $E$, construct the matrices 

\begin{align}
  \begin{split}
    M_1(E) &= E \, I - H_0 + i \epsilon \, I \\
    M_2(E) &= E \, I - H + i \epsilon \, I;
  \end{split}
\end{align}

 and invert

\begin{align}
  \begin{split}
    \hat{G}_0(E) &= \{ M_1(E)\}^{-1} \\
    \hat{G}(E) &= \{ M_2(E)\}^{-1}.
  \end{split}
\end{align}

\item[(iii)] Extract the quantities of interest. For example, the propagators are obtained from

\begin{align}
  \begin{split}
    G_\rho &= \hat{G}(1,1) \\
    G^0_\rho &= \hat{G}_0(1,1) \\
    G_{2\pi}(q_i) &= \hat{G}(i,i) \\
    G^0_{2\pi}(q_i) &= \hat{G}_0(i,i).
  \end{split}
\end{align}

\noindent where $q_i$ is the discrete momentum of the $i$-th grid.

\item[(iv)] The spectral functions $A$'s can be obtained from the propagators by simply taking the imaginary part.
  To calculate the $B$ function, one possible method is

\begin{align}
  \begin{split}
  B(E) &= -2 \, {\rm Im} \, {\rm tr} \, \left( G(E) - G_0(E) \right) \\
  &+ 2 \, {\rm Im} \, \hat{G}_0(1,1). 
  \end{split}
\end{align}

\end{itemize}

In this exercise, we have chosen an appropriate P-wave coupling to describe the physical $\rho$-meson:

\begin{align}
g_{\rm eff} = g \times q \times e^{-\frac{q^2}{2 \Lambda^2}},
\end{align}

\noindent with parameters $g = 23.5 \, {\rm GeV}^{-3/2}$, $\Delta_0 = 0.64 $ GeV, and $\Lambda = 0.4$ GeV. 
The form factor renders the real part of the self-energy finite.~\footnote{The imaginary part is finite even without a regulator. We checked that its value is only slightly modified by the form factor.} It can be motivated in a quantum field theory 
with non-local interactions or dressing of vertex~\cite{nonlocal1,nonlocal2,nonlocal3,nonlocal4}. 
For what concerns us here it parametrizes the finite-size effects of the hadrons. 
See Ref.~\cite{Herrmann:1993za} for a rigorous treatment within the QFT framework.

The numerical results for $\mathcal{Q}$ and various spectral functions are shown in Fig.~\ref{fig:one}. 
We have computed the phase shift on the momentum grid, ($1+N_q=800$; $L\approx800$ fm; $\epsilon=0.01$ GeV), in two ways:
one uses $G_\rho$ via Eq.~\eqref{eq:ps1} (grey squares), 
the other uses the scattering matrix $S$ via Eq.~\eqref{eq:recipe} (black circles). 
Both results agree quite well with the continuum limit, 
although they appear to have different convergence property.~\footnote{ 
Within this model one can also study how the phase shift 
and the spectral functions depend on the lattice size. 
This may help in understanding phase shift extraction in LQCD~\cite{FVH,LQCD}.}

A key feature to note is the apparent shift of the strength of $B$, compared to $A_\rho$, towards lower energies.
This effect originates from the nonresonant scattering term $\sum_q \Delta A_{2 \pi}(E;q)$, 
and is needed, in addition to $A_\rho$, for a complete description of the interacting system.

\subsection{Phase shift from S-matrix}

In Eq.~\eqref{eq:ps1} we extracted the phase shift $\mathcal{Q}$ from the $\rho$-propagator. 
The same information is also available from the 2-pion sector via the S-matrix.
In Ref.~\cite{smat}, the following recipe has been proposed to extract the phase shift from an $N_q \times N_q$ S-matrix

\begin{align}
  \label{eq:recipe}
  \begin{split}
    \mathcal{Q} &= \frac{1}{2} \times {\rm Im} \, {\rm tr} \,  \ln S \\
                &= \frac{1}{2} \times {\rm Im} \, \ln {\rm det} S \\
                &\approx  \frac{1}{2} \times {\rm Im} \, \ln (1 + i \times  {\rm tr} \, \hat{t}).
  \end{split}
\end{align}

\noindent where we have introduced the $\hat{t}$ operator, defined as

\begin{align}
  \hat{t} = - 2 \pi \, \delta(E-H_0) \times T.
\end{align}

\noindent From Eqs.~\eqref{eq:sigma} and~\eqref{eq:tmat} we see

\begin{align}
\begin{split}
  {\rm tr} \, \hat{t} &= -2 \, \sum_q  \left( \pi \, \delta(E-\epsilon(q)) \times  g_{\rm eff}^2 C(q)^2 \right) \times G_\rho \\
  &= 2 \, {\rm Im} \, \Sigma_\rho \times G_\rho, 
  \end{split}
\end{align}

\noindent and indeed it is straightforward to verify the same result of the phase shift as in Eq.~\eqref{eq:ps1}:

\begin{align}
  \begin{split}
    &\frac{1}{2} \times {\rm Im} \, \ln (1 + i \times  {\rm tr} \, \hat{t}) \\
    &= \frac{1}{2} \times {\rm Im} \, \ln \left( {G_\rho^{-1}}^\star / \, G_\rho^{-1} \right) \\
    &= \tan^{-1} \, \frac{{\rm Im} \, \Sigma_\rho}{E-\Delta_0-{\rm Re} \Sigma_\rho}.
  \end{split}
\end{align}

The approximation in the last line of Eq.~\eqref{eq:recipe} was shown to be valid for some simple cases such as s-channel-only interaction or structureless scattering. We shall now show that the approximation is exact in the Lee model. 

Consider the expansion of the logarithm of the S-matrix via the Mercator series

\begin{align}
  \begin{split}
    {\rm tr} \ln S &= {\rm tr} \, \ln (I + i \times \,  \hat{t}) \\
                   &= \sum_{m=1}^\infty \frac{(-1)^{m+1} i^{m} }{m} \, {\rm tr}\, ({\hat{t}}^m).
  \end{split}
\end{align}

\noindent The approximation in Eq.~\eqref{eq:recipe} becomes exact if 
$\hat{t}$ satisfies the following property

\begin{align}
  \label{eq:criteria}
  {\rm tr}\, ({\hat{t}}^m) \rightarrow ({\rm tr} \, \hat{t})^m,
\end{align}

\noindent giving

\begin{align}
  \begin{split}
    \sum_{m=1}^\infty \frac{(-1)^{m+1} i^{m} }{m} \, {\rm tr}\, ({\hat{t}}^m)  &\rightarrow \sum_{m=1}^\infty \frac{(-1)^{m+1} i^{m} }{m} \, ({\rm tr}\, {\hat{t}})^m \\
                   &=  \ln (1 + i \times  {\rm tr} \,  \hat{t}).
  \end{split}
\end{align}

\noindent Note the effective replacement of $I$ with $1$ in the above.

Inspecting the interaction term of the Lee Model in Eq.~\eqref{eq:vint}, 
we see that the matrix element of $\hat{t}$ takes the product form

\begin{align}
  \hat{t}_{q_1 q_2} = u_{q_1}v_{q_2}.
\end{align}

\noindent It follows that

\begin{align}
  \begin{split}
    ({\hat{t}^2})_{q_1 q_2} &= \sum_{q_3} u_{q_1}v_{q_3} u_{q_3}v_{q_2} \\
                            &= u_{q_1}v_{q_2} {\rm tr} \, \hat{t}, 
  \end{split}
\end{align}

\noindent and hence $ {\rm tr}\, ({\hat{t}}^2) = ({\rm tr} \, \hat{t})^2 $ and so on for higher power.
This demonstrates that criteria~\eqref{eq:criteria} is satisfied by
a class of $\hat{t}$ which involves separable potentials, i.e. the ones that are from a Kronecker (direct) product.

Furthermore, we add that from Cayley Hamilton theorem, the determinant of $\hat{t}$ essentially vanishes. 
In fact, all eigenvalues of $\hat{t}$ are zero, except one, which is ${\rm tr} \, \hat{t}$. 
This is another way to understand why the approximation made in Eq.~\eqref{eq:recipe} is justified.
The full implication of this result is not yet completely clear, and will be explored in a future work.

\section{Thermodynamics}
\label{sec4}

The change in the density of state due to interactions, as revealed by the KFL $\mathcal{K}$ operator or the $B$ function, is the key input for the S-matrix formulation of statistical mechanics~\cite{dmb,venugopalan}. The approach is based on the method of cluster expansions, and for the second virial coefficient the result is exact. We retrace a few basic steps in relating the scattering phase shift to the thermal partition function. 

Our starting point is the cluster expansion of the grand partition function

\begin{align}
  \begin{split}
  Z(T, V, \xi) &= {\rm tr} \, e^{-\beta (H-\mu N)} \\
               &= \sum_N  Z_N \xi^N
  \end{split}
\end{align}

\noindent where $\xi$ is the fugacity, related to the particle chemical potential via $\xi=e^{\mu/T}$. 
$Z_N$ is the N-body partition function. The corresponding expansion for the logarithm of $Z$ reads

\begin{align}
  W(T, V, \xi) = \ln Z = \sum_N  W_N \, \xi^N, 
\end{align}
\begin{figure*}[ht!]
\centering
\includegraphics[width=0.497\textwidth]{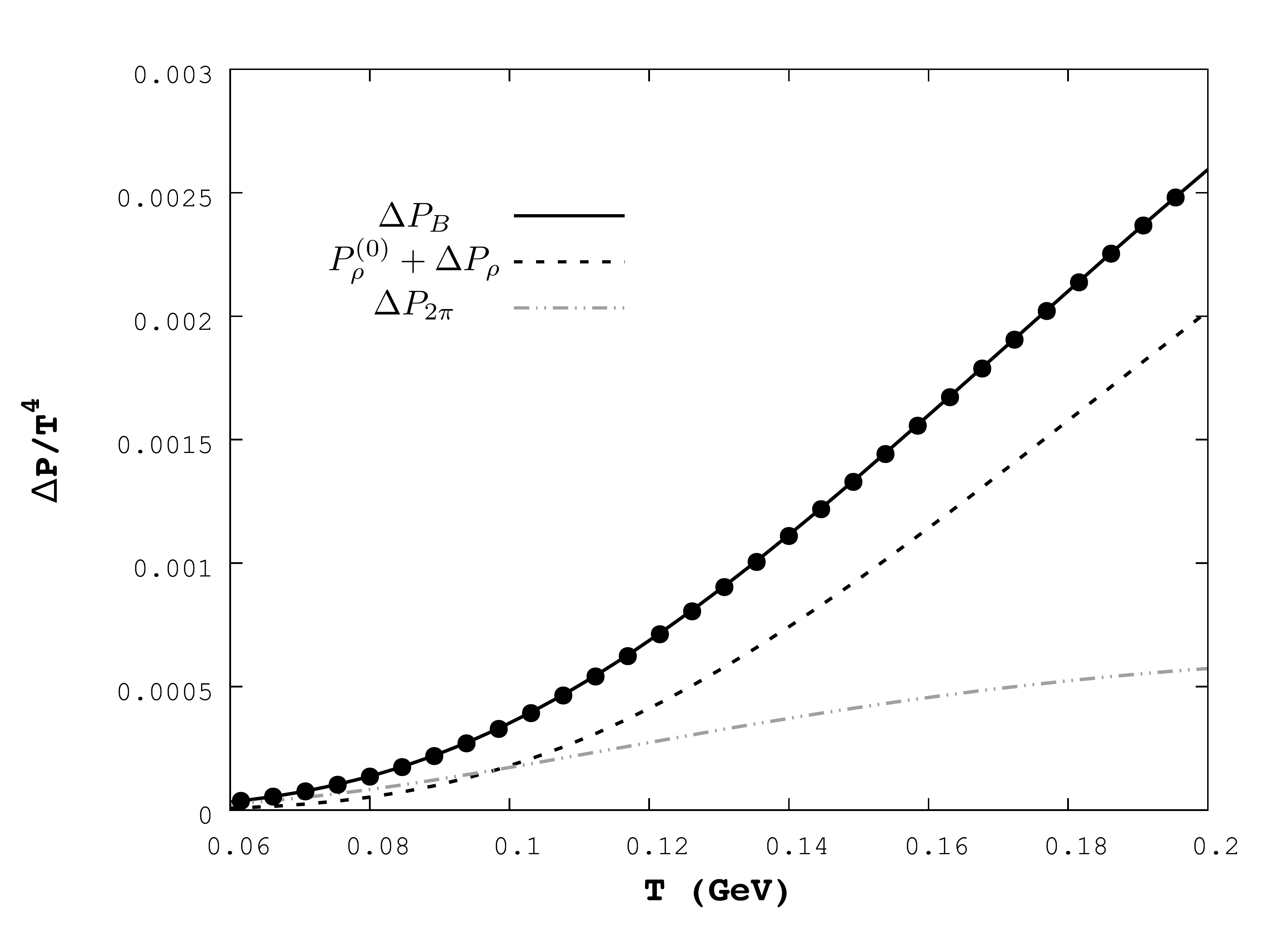}
  \caption{Interacting contributions to the thermodynamics pressure (normalized to $T^4$) from different effective spectral functions. The top line ($B$) is equal to the sum of the contributions from the lower two. (see Eq.~\eqref{eq:pressure}) 
  The points are the results obtained by directly constructing the partition function from the eigenvalues of the Hamiltonian. See text.
  }
\label{fig:two}
\end{figure*}

\noindent and one can work out

\begin{align}
  \begin{split}
    W_1 &= Z_1 \\
    W_2 &= Z_2 - \frac{1}{2} Z_1^2 \\
    W_3 &= Z_3 -Z_1 Z_2 - \frac{1}{3} Z_1^3 \\
    \cdots.
  \end{split}
\end{align}

\noindent The interacting part of the partition function satisfies~\footnote{For simplicity we neglect corrections due to quantum statistics and focus on the $\mu=0$ case.}

\begin{align}
  W_2 - W_2^{(0)} = Z_2 - Z_2^{(0)},
\end{align}

\noindent The latter can be re-expressed via
\begin{align}
  \label{eq:z2}
  \begin{split}
  Z_2 - Z_2^{(0)} &= {\rm Tr} (e^{-\beta H_2}-e^{-\beta H_2^{(0)}}) \\
                  &= V \int \frac{d^3 P}{(2 \pi)^3} \frac{d E}{2 \pi} \, e^{-\beta E_{\rm sys}(\vec{P}, q)} \, D(E) \\
                  &= \beta V \times (T \lambda_{T}^3) \,  e^{-\beta m_{\rm tot}} \times \int \frac{d E}{2 \pi} \, e^{-\beta E}\, D(E).
  \end{split}
\end{align}

\noindent where we have integrated out the CM motion in the total energy of the 2-body system

\begin{align}
    E_{\rm sys}(\vec{P}, q) &= m_{\rm tot} + \frac{\vec{P}^2}{2 m_{\rm tot}} + E,
\end{align}
  
\noindent with $m_{\rm tot}$ being the total mass. $E$ is the energy of the relative motion

\begin{align}
  E = \frac{q^2}{ 2 m_{\rm red}}
\end{align}

\noindent where $m_{\rm red}$ is the reduced mass. 
We have made use of the form of the nonrelativistic dispersion (see e.g., Ref.~\cite{smat} for a relativistic formulation) to integrate out the CM momentum $\vec{P}$ 
and obtain the thermal wavelength $\lambda_T$

\begin{align}
  \lambda_{T}^3 \equiv  \int \frac{d^3 P}{(2 \pi)^3} \, e^{-\beta \frac{\vec{P}^2}{2 m_{\rm tot}}}. 
\end{align}

\noindent Note that the bare $\rho$ state is not counted in the trace of $Z_2^{(0)}$. 
The remaining integral in Eq.~\eqref{eq:z2} requires the input of $D(E)$, 
which is the change in the density of state due to interactions. 

Now we are ready to examine the thermal contribution of an unstable state $\rho$
based on the input from the Lee model.
The first thing to notice is that the L.H.S. of Eq.~\eqref{eq:z2} 
can be directly computed from the eigenvalues ($\lambda_n$) of the Hamiltonian:~\footnote{In fact, computations involving the difference between the fully interacting system and the free case, e.g., $G(z)-G_0(z)$, are numerically more stable~\cite{BALIAN1970401}. This also opens up the possibility of solving the system with other numerical techniques, such as the use of a harmonic oscillator basis in the expansion~\cite{Kruppa:1998pvy}.}

\begin{align}
\begin{split}
{\rm Tr} \, e^{-\beta H} &= \sum_{\lambda_n} \, e^{-\beta \lambda_n} \\
  &= -2 \, {\rm Im} \, \int \frac{d z}{2 \pi} \, e^{-\beta z} \, {\rm Tr} \, G(z). 
\end{split}
\end{align}

\noindent The result is shown as points in Fig.~\ref{fig:two}.
This in turns indicates what is the correct $D(E)$ to use.
The S-matrix formulation of statistical mechanics by Dashen {\it et al.}~\cite{dmb} 
dictates the choice of $D(E) \rightarrow B(E) = 2 \frac{\partial}{\partial E} \mathcal{Q}(E)$. 
This means that we compute the thermodynamic pressure via

\begin{align}
\label{eq:p1}
  P_{\rm S-matrix} \approx P^{(0)}_{\pi} + \Delta P_B.
\end{align}

\noindent $\Delta P_B$ includes the contribution of the unstable state $\vert \rho \rangle$ and 
nonresonant $\pi \pi$ interaction: 

\begin{align}
    \Delta P_{B} &= P_{\rho} + \Delta P_{2 \pi}.
\end{align}

From Eq.~\eqref{eq:golden} and the discussion it is clear that one can incorporate the same physical content of the thermal medium with 
a different choice of $D(E)$, though with a different interpretation. 
For example, one can choose instead $D(E) \rightarrow {\rm tr} \mathcal{K}$, and in this case 
the contribution of $A^0_\rho$ needs to be added separately as

\begin{align}
\label{eq:p2}
  \begin{split}
    P_{\rm KFL} &\approx P^{(0)}_\pi + P^{(0)}_{\rho} + \Delta P_{\mathcal{K}} \\
    &= P^{(0)}_{\pi} + P^{(0)}_{\rho} + \Delta P_{\rho} +  \Delta P_{2 \pi}.
  \end{split}
\end{align}

\noindent Here $\Delta P_{\mathcal{K}}$ contains the contribution from $\Delta A_\rho$ and $\Delta A_{2 \pi}$. 
Note that as $g \rightarrow 0$, $\Delta P_{\mathcal{K}} \rightarrow 0$; while $\Delta P_B \rightarrow P^{(0)}_\rho$.
The two results in Eqs.~\eqref{eq:p1} and~\eqref{eq:p2} are equivalent, i.e.

\begin{align}
\label{eq:pressure}
  \begin{split}
    \Delta P_{B} &= P^{(0)}_{\rho} + \Delta P_{\mathcal{K}} \\
    &=  P^{(0)}_{\rho} + \Delta P_{\rho} +  \Delta P_{2 \pi}.
    \end{split}
\end{align}

\noindent which is just restating the relation~\eqref{eq:golden}. 

The various partial pressures are shown in Fig.~\ref{fig:two}.
Due to the contribution from the nonresonant $\pi \pi$ state, 
the pressure based on $B$ is substantially larger than the one based on $A_\rho$ alone.
We stress that only the former one gives a consistent description of the thermodynamics,
as can be verified by the direct construction of the partition function from the 
eigenvalues of the Hamiltonian (black circles).
Therefore Eq.~\eqref{eq:pB} should be used instead of Eq.~\eqref{eq:pA}.

These observations are in accord with the previous analysis based on a $\Phi$-derivable approach~\cite{weinhold}. 
The $B$-function-based description requires only the input from the scattering of asymptotic states. 
This underlines an important concept in the formulation:
In computing the density of states it is not mandatory to introduce the unstable state as an explicit degree of freedom.
For approaches that use only stable states as degrees of freedom, 
such as an effective field theory where resonances are dynamically generated~\cite{Kaplan:1996xu}, 
the same density of states would be obtained as long as the phase shifts agree.
And when an empirical phase shift $\mathcal{Q}(E)$ is used, the function $B(E)$ becomes model independent, 
while the splitting into $A_\rho(E)$ and $\sum_q \Delta A_{2 \pi}(E;q)$ is model dependent.
The Lee model studied here provides a clear picture of such a splitting, 
and demonstrates how an unstable state should be included in the description of the thermodynamics. 

\section{Conclusion}
\label{sec5}

In the context of the Lee model we have clarified the relation between 
the energy derivative of the phase shift and 
the spectral functions of the degrees of freedom composing the system.
We have also illustrated how these quantities enter the thermal description of the system via the 
S-matrix formulation of the statistical mechanics.
This consolidates our understanding of the connection between this and the standard approach 
based on thermal Green functions. In particular, we have shown that 
the thermodynamic trace requires the inclusion of the nonresonant contribution ($\Delta A_{2 \pi}$), 
in addition to, and independent of, the effect coming from the width of the unstable state $\Delta A_{\rho}$.

Besides acting as an effective density of state, an alternative interpretation of the energy derivative of the phase shift is the concept of time delay~\cite{Danielewicz:1995ay,
Kelkar:2008na}: particles spend longer or shorter in the interaction region due to the attractive or repulsive nature of the interaction. 
In the contexts of transport models and resonance identification, 
it was argued~\cite{bass,leupold1,leupold2,Kelkar:2003iv,KELKAR2004121} that such a time delay, instead of the inverse width $1/\Gamma(E)$, should be used to measure the life-time of a resonance. 
A related problem is the study of the survival probability of an unstable state. According to Ersak~\cite{ersak,Fonda:1978dk}, the standard exponential decay law is valid only in the limited case of an energy-independent Breit-Wigner distribution. 
Re-scattering effect, apparently related to $\Delta A_{2 \pi}$, would lead to non-exponential behavior~\cite{Bohm:2004zi}. 
A clearer theoretical understanding of $B$ and $A_\rho$ could provide further insights into these topics.

So far we have restricted our discussion to Fock space up to two body. 
It will be extremely interesting to extend the scheme to include 
multi-channel and multi-body scatterings~\cite{Chiu:1992pc,Kaminski:1996da,Mai:2017bge,smat,chibs}, 
and understand how these interactions would influence thermodynamic quantities. 
This can be a useful framework to analyze the observables in Heavy Ion Collision experiments, 
such as hadron yields and the momentum distributions of light hadrons~\cite{LHC,Sollfrank:1990qz,Broniowski_2003,omega}.
We defer this more challenging problem to future research.



\section*{acknowledgments}

PML thanks Eric Swanson for stimulating discussions.
He also acknowledges fruitful discussions with Hans Feldmeier, Bengt Friman, and Piotr Bozek.
The authors are also grateful for the constructive conversations with Wojciech Broniowski, Wojciech Florkowski, and Stanislaw Mrowczynski. 
PML was partly supported by the Polish National Science Center (NCN), under Maestro Grant No. DEC-2013/10/A/ST2/00106 
and by the Short Term Scientific Mission (STSM) program under COST Action CA15213 (reference number: 41977). 
FG acknowledges financial support from the Polish National Science Centre (NCN) through the OPUS project no. 2015/17/B/ST2/01625.  

\bibliography{ref}

\end{document}